\documentstyle[12pt,epsf]{article}
\newcommand{\be}{\begin{equation}}
\newcommand{\ee}{\end{equation}}
\newcommand{\bea}{\begin{eqnarray}}
\newcommand{\eea}{\end{eqnarray}}
\newcommand{\rf}[1]{(\ref{#1})}
\newcommand{\nn}{\nonumber}

\def\boxit#1{\vbox{\hrule\hbox{\vrule\kern3pt\vbox{\kern3pt#1\kern3pt}\kern3pt\vrule}\hrule}}
\newdimen\hsbox
\hsbox=\hsize \advance\hsbox by 0.28truein
\def\DefWarn#1{}
\def\del{\partial}
 \def\Tr{{\rm Tr}}
 \def\bar{\overline} %% \def\hat{\widehat}

 \def\CS{{\cal S}} \def\CT{{\cal T}}
\def\vx{{\bf x}}

\def\frac#1#2{{{#1}\over {#2}}}

\def\third{\hbox{${1\over 3}$}}

\def\PR#1#2#3{{ Phys.~Rev.~}{ #1} (#2) #3}
\def\PRL#1#2#3{{ Phys.~Rev.~Lett.~}{ #1} (#2) #3}
\def\NP#1#2#3{{ Nucl.~Phys.~}{ #1} (#2) #3}
\def\PL#1#2#3{{ Phys.~Lett.~}{ #1} (#2) #3}

\def\AP#1#2#3{{ Ann.~Phys.~}{ #1} (#2) #3}

\def\JP#1#2#3{{ J.~Phys.~}{ #1} (#2) #3}

\def\vol#1{{ #1}}
\def\vyp#1#2#3{\vol{#1} (#2) #3}

\begin{document}
\topmargin 0pt
\oddsidemargin 5mm
\headheight 0pt
\topskip 0mm

\newdimen\hsgraph \newdimen\vsgraph
\hsgraph=\hsize \advance\hsgraph by-1.2truein
\vsgraph=\vsize \advance\vsgraph by-1.5truein

\pagestyle{empty}

\begin{flushright}
OUTP-96-23P\\
1st July 1996\\
hep-lat/9607005
\end{flushright}

\begin{center}

\vspace{18pt}
{\Large \bf Critical Properties of the Z(3) Interface in (2+1)-D SU(3) Gauge Theory}

\vspace{2 truecm}

{\sc S.T. West\footnote{e-mail: west@thphys.ox.ac.uk} and
 J. F. Wheater\footnote{e-mail: jfw@thphys.ox.ac.uk} }

\bigskip
{\it 
Department of Physics, University of Oxford, \\
Theoretical Physics,\\
1 Keble Road,\\ Oxford OX1 3NP, UK}
\vspace{3 truecm}

\end{center}

\noindent
{\bf Abstract.} We study the interface between two different $Z(3)$ vacua
in the deconfined phase of $SU(3)$ pure gauge theory in $2+1$ dimensions
just above the critical temperature. In simulations of the Euclidean
lattice gauge theory formulation of the system we measure the fluctuations 
of the interface as the critical temperature is approached and as a function
of system size.  We show that the intrinsic width of the interface remains
small even very close to the critical temperature. Some dynamical exponents
which govern the interaction of the interface with our Monte Carlo algorithm
are also estimated. We conclude that the $Z(3)$ interface has properties 
broadly similar to those in many other comparable statistical mechanical systems.

\vfill
\newpage
\pagestyle{plain}

\section{Introduction}
It has been known for some time that pure gauge theories have a
non-confining high-temperature phase \cite{IRoo}and so  any 
theory which is in a
confining phase at zero temperature ($T=0$) must have a 
deconfining phase transition at
some finite critical temperature, $T_c$.
Pure non-abelian $SU(N)$ gauge theories, such as QCD without
 dynamical fermions,
display this behaviour in both $2+1$ and $3+1$ dimensions.
 At low temperatures the
colour charge of QCD is confined which
 is why no
free quarks are seen in the relatively cool universe of today.
At high temperatures, there is a
non-confining phase, corresponding to
the free quark-gluon plasma believed to exist in the hot early
universe.
In fact there are in general several degenerate phases of this
type -- as many as the number of elements of the centre, $Z(N)$, of the
gauge group.
 Fermionic matter breaks the vacuum degeneracy, albeit on a
small scale, so we consider only the pure gauge theory in what follows.

Two different types of interface are possible
in the theory. The first is  between the ordered and
disordered phases and  is only stable at the critical
temperature where the two phases can coexist. Secondly, an interface can 
form between two of the
ordered phases with different $Z(N)$ vacua, and this is usually called
an ``order-order interface'', or ``$Z(N)$ interface''.  Clearly,
this  type of interface  can only exist above the critical
temperature.

From the thermodynamic point of view there is something strange about
these interfaces. On entropic grounds one expects to find disorder at high
temperatures and order at low temperatures, not the other way round.
The fundamentally Euclidean nature of this picture may be 
responsible  \cite{IRoba,IRoc,IRod}; there is no sensible counterpart to
the Euclidean order parameter in Minkowski space (a naive Wick rotation 
leads to an imaginary time-like gauge field in Minkowski space)
and so it may be that the interfaces do not exists as physical objects 
in the high temperature universe.
Indeed, it has even been  suggested  \cite{IRoe} that only one true physical
phase exists, even in Euclidean space, at high temperatures but recent
numerical work does not support this point of view.
Whether or not they are a Euclidean artifact,
by virtue of their contribution to the partition function
and to expectation values calculated using the Euclidean path
integral, the interfaces must be included (like the instanton) in a
non-perturbative analysis of the thermodynamics of QCD.

Recently the properties of $Z(N)$ interfaces at very high temperatures
have been investigated numerically in $2+1$ dimensions
\cite{IRxiia,stwjfw}.
These simulations give a 
non-perturbative check on the conclusions of analytic calculations
\cite{IRx,IRix}
and show very good agreement; thus there is no reason to suppose that the 
$Z(N)$ phases are not distinct at high temperatures. In this paper we
investigate the behaviour of the interface at much lower temperatures
close to the critical
temperature where it disappears. We shall be concerned not with the 
usual thermodynamical quantities, which have been thoroughly investigated
\cite{IIIRi}, but rather with the geometrical properties and structure of
the interface itself.

This paper is organised as follows. In section two we review briefly
the theory of the deconfinement transition and the theoretical basis on 
which we shall analyse the simulation data on the interfaces.  In section 
three we discuss the practical difficulties in identifying the interface
from the gauge field configuration and describe how we have overcome them;
some other aspects of the data analysis are also discussed. Section four
contains the results that we have obtained characterising the interface
fluctuations, and section five gives our results for various other properties
of the interface. In section six, we
give our conclusions and discuss some of the remaining puzzles associated
with these interfaces.

\section{Theoretical Background}
In the Euclidean formalism, equilibrium finite-temperature field
theory is obtained
by considering a system which is of infinite volume but compact in the 
Euclidean time direction with period $\beta_T=T^{-1}$.
In a lattice field theory implementation, it is of course not possible
to vary the 
number of lattice spacings in the time direction, $L_t$, continuously;
instead we 
work with a fixed $L_t$ and vary the gauge coupling, $\beta$, so that
the lattice
correlation length, and hence the physical value of the lattice
spacing, $a$, and hence the physical $\beta_T=L_ta$, vary continuously.

The phases of a  gauge theory at finite temperature are characterised
by the free energies of static configurations of quarks and
antiquarks \cite{IRv}.
The self-energy of a single quark in the gluon medium, $F_q$,
is related to  the expectation value, $L(\vx)$, of a single Polyakov line, 
a time-ordered Wilson loop which wraps
around the time boundary for a fixed spatial location $\vx$,
\be{L(\vx)=\third\Tr\CT e^{ig\int_0^{1/T} d\tau
A_0(\tau,\vx)}=e^{-{\beta_T} F_q}.}\ee
If $<L(\vx)>=0$ then the insertion of a single quark will require
infinite energy, corresponding to a confining phase. In contrast, if
$<L(\vx)>\neq 0$ then the insertion energy will be finite and isolated colour
charges can exist -- this is the deconfined phase. 
Thus, the expectation value of the Polyakov line gives an effective test for
confinement.

A general non-Abelian gauge transformation which leaves the Euclidean
QCD action, $\CS_E$, invariant will also leave $L$ invariant if the
transformation is periodic in time. However, 
$\CS_E$ is actually invariant under an additional global symmetry. In the 
lattice formulation this consists of multiplying each time-like link variable
at a given time slice by an element of the
centre of the gauge group (the set of
elements which commute with all members of the group, $Z(N)$ in the
case of $SU(N)$). This global symmetry
is an invariance of $\CS_E$ and cannot be undone by a 
 gauge transformation.
The topologically non-trivial
Polyakov line, which wraps around the periodic boundary condition in
the time direction, is not invariant under these transformations, but 
is rotated by an element, $z$, of the centre: $L(\vx) \rightarrow z L(\vx).$
Clearly, $<L(\vx)>$ acts as an order parameter for the centre
symmetry, distinguishing between broken and unbroken phases, and
between the different broken phases \cite{IRx,IRviioa}.

In the deconfined phase, where the $Z(N)$ symmetry has been spontaneously
broken, the system can exist in any one of $N$
degenerate vacua. It is possible to arrange boundary conditions
of the system so that different parts of it exist in different
vacua, {\it i.e.} different $Z(N)$ phases. The existence of these
distinct domains forces the appearance of domain walls, or ``$Z(N)$
interfaces'', where they meet \cite{IRx,IRix}.
Within these interfaces, the gauge
fields interpolate between the different vacua, as does the
expectation value of the Polyakov line. A perturbative calculation
at large $\beta$ \cite{IRx} leads to the conclusion that an 
instanton interpolating between two vacua should have characteristic
size, $l_W$ of order $1/g\sqrt{T}$; therefore, at least at large enough 
$\beta$, we expect the interface to have an intrinsic width of this
magnitude as well. This length scale is the same order as the 
Debye screening length, $l_D$ which
governs gauge-invariant correlation functions of the time-like
component of the gluon field ($A_0$) at large distances and high
temperatures. The free energy of a quark-antiquark
pair, over and above the sum of their separate free energies,
vanishes as the quarks become infinitely far apart,
and, from perturbation theory, we know that quarks are screened at a
distance of the order of the inverse electric mass, so rather than
being logarithmic in $\vx$, the interaction potential takes
the form
\be F_{q\bar
q}(\vx)-2F_q=V(|\vx|,T)\mathrel{\mathop{\sim}\limits_{|\vx|\rightarrow
\infty}} -C\exp\left(-2{|\vx|\over l_D}\right) \qquad\hbox{for\ } T>T_c.\ee
The factor of two arises because gauge invariance leads to an exchange of
two gluons being the lowest-order contribution in perturbation theory.
The Debye screening length is known \cite{dHoker} to have logarithmic
corrections so that
\be l_D={1\over g\sqrt T}.
{1\over {3\over 4\pi}\log({\beta\over 12})+\ldots}\ee
The corresponding corrections to $l_W$ have not been calculated so,
even at very high temperature, a direct comparison between the 
interface width and the Debye screening length is not possible. 
At low temperatures, near the critical point, we expect all these
quantities to be functions of $\beta-\beta_c$ and there are no 
analytic results available with which to compare our simulations.

At very high temperatures the magnitude of $<L>$ is large; the change
in its argument 
from one $Z(N)$ phase to the other is abrupt. The interface between
the two phases
is sharp (see fig. 2 of section 3); it is essentially a
one-dimensional object that separates the phases
and exhibits relatively small transverse fluctuations. As $\beta$
decreases towards its critical value, $<L>$ vanishes according to 
\be<L>\sim (\beta-\beta_c)^{\beta_M},\ee
where $\beta_c = 8.175(2)$ in the infinite spatial volume limit on
$N_t=2$ lattices, and the measured critical exponent $\beta_M$
\cite{IRxv,IIIRi}
is consistent
with the universality prediction of ${1\over 9}$ from the three-state
Potts model \cite{IRvii}.
However, this only tells us about the expected height of the interface
and not about its geometrical nature as $\beta$
decreases (except that it must ultimately disappear at $\beta_c$). It
may be that,
as the free energy per unit length declines, the interface undergoes
increasingly violent fluctuations while remaining intrinsically a
one-dimensional object, the height of the interface
declining until it vanishes at $\beta_c$. An  alternative is that  the interface
may become 
intrinsically broader, thus becoming more two-dimensional, with more
and more regions of
disordered phase within it. At $\beta_c$ these disordered regions
expand to take over the whole
system and the interface disappears.

\begin{figure}
{\epsfxsize=\textwidth \epsfbox{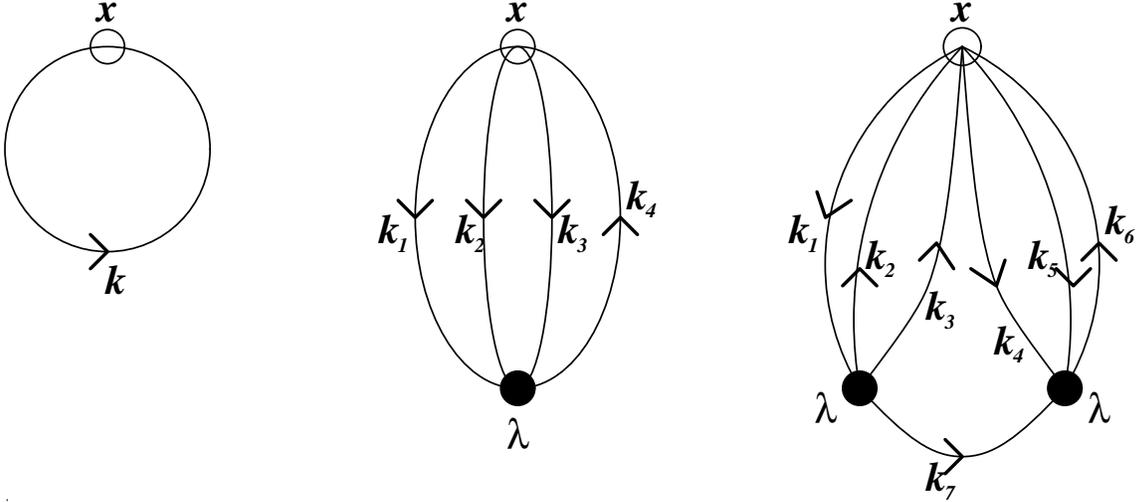}}
\caption{The vacuum diagrams corresponding to the 2nd, 4th and 6th
connected vacuum correlation functions respectively. The open circle
simply denotes point $\vx$, the endpoint of each external leg; the
solid circle denotes an interaction vertex, $\lambda$; and the
momentum travelling along a leg is denoted by $k$.}
\end{figure}

To help in the analysis of the fluctuations of the interface we shall
compare them to a simple model. 
Our lattice is taken to have $L_x\sim3L_y\gg \L_t$ ; the boundary
conditions are periodic and a 
``twist'' is introduced (after \cite{IIRo}) to ensure that for $x\sim
0$ the system is in one $Z(3)$ phase while
for $x\sim L_x$ it is in a different one. The interface thus forms
stretching across the lattice in the 
$y$ direction and separating the two phases.  Now let $\phi(y)$ denote
the displacement, in the $x$ direction, of the 
interface from its equilibrium position and construct a simple
effective Lagrangian, $\cal L$, for $\phi$. 
Translational invariance dictates that $\cal L$ can only depend on
$\partial_y\phi$ and invariance 
under $\phi\to-\phi$ rules out odd powers; if the interface were just executing
a free random walk across the lattice then ${\cal L}\sim (\del_y\phi)^2$.
To obtain non-gaussian moments a correction term must be added and
the simplest resulting Lagrangian is
\be{\cal L}={1\over 2}\gamma(\partial_y\phi)^2+{\lambda\over
4!}(\partial_y\phi)^4\ee
which
fits the data quite well with 
\bea \gamma&\sim&(\beta-\beta_c)^{\nu_1}\nn\\
 \lambda&=& \ell\lambda_0\nn\\
\lambda_0&\sim&(\beta-\beta_c)^{\nu_2}\eea
where the length scale $\ell$ appears in $\lambda$ on dimensional grounds.
It can be any combination of the two length scales which may be relevant,
$L_y$ and $l_D$.
The diagrams for the first three even
moments of $\phi$ are shown in fig.1.
It is straightforward to compute the connected correlation functions
from $\cal L$ although
the mode sums have to be done numerically. Putting $\ell=L_y$
the connected correlation
functions (denoted by $<.>_c$) are then predicted to behave like
\bea <\phi^2>&\sim& L_y\gamma^{-1}\nn\\
<\phi^4>_c&\sim& L_y^2\gamma^{-4}\lambda_0\nn\\
<\phi^6>_c&\sim& L_y^3\gamma^{-7}\lambda_0^2\label{moments}\eea

\section{Identifying the Interface}
Fig.2 shows the $Z(3)$ interface on a particular gauge field
configuration for a number of $\beta$ 
values. As discussed in section 2, at high $\beta$ the interface is
sharp and well defined; however  
by $\beta=9$, still some way above the critical value, it is becoming
difficult to identify the location 
of the interface by eye with any degree of certainty.

\begin{figure}
{\epsfxsize\textwidth \epsfbox{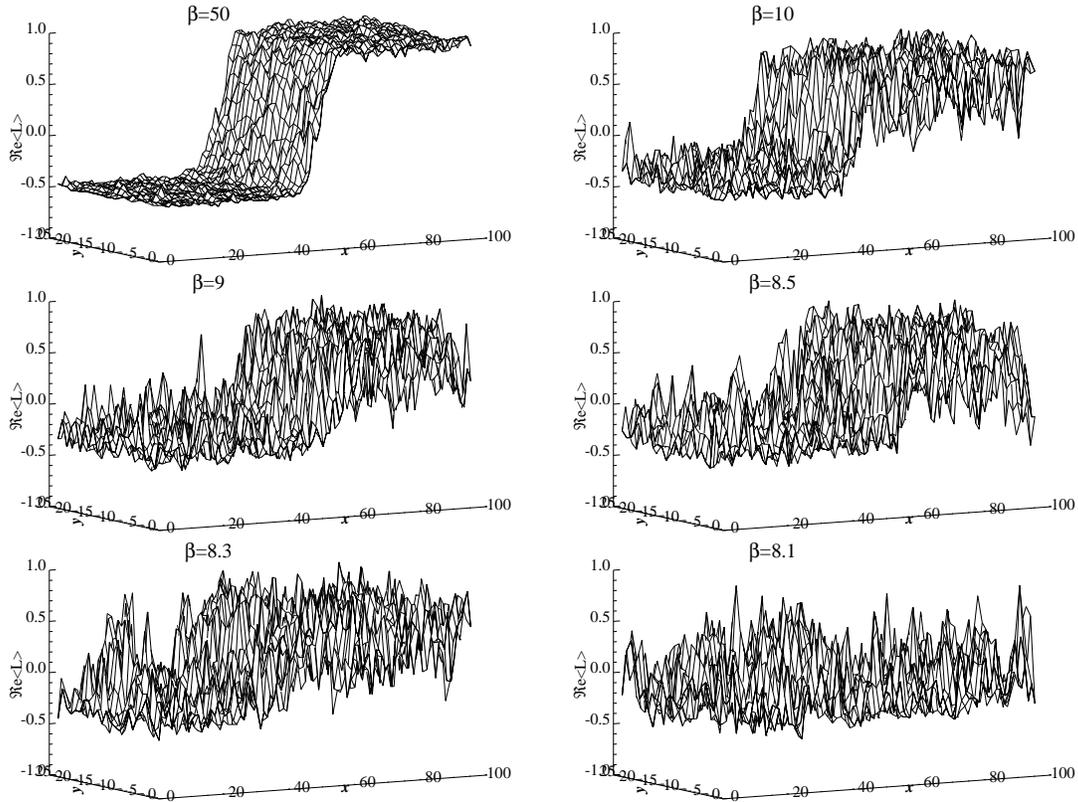}}
\caption{Polyakov line (real part) profiles of the same $Z(3)$ interface at
different temperatures. The top left picture shows the interface on
a $2\times 24\times 96$ lattice with $\beta=50$. Going
from left to right, and then top to
bottom, the other pictures illustrate an interface between the same
phases ($<L>\approx e^{2\pi i/3}$ on
the left and the $<L>\approx 1$ on the right) for $\beta=10, 9, 8.5, 
8.3$ and $8.1$ respectively.}
\end{figure}

One way to deal with the increase in phase fluctuations is to cut out
the highest frequency modes altogether. This can be achieved most
simply by a
``box-car'' average, where the Polyakov lines of each three-by-three
spatial array of points
are averaged over to give a new value, allocated to the central
point. 
The effect of smoothing out the fluctuations is shown in
fig.3, produced by applying the
box-car average to the pictures of fig.2.
To make things even clearer, the Polyakov lines
have first been re-binned so that the $24\times 96$ spatial dimensions
become $12\times 48$, with each two-by-two array being averaged to
one point. Now the $Z(3)$ interface can clearly be seen for all
but the last picture, {\it i.e.} for all temperatures above
$\beta_c$. The fact that the interface can still be seen for
temperatures just above the critical value suggests
that a study of its structure should still be possible quite close to
the phase transition.

\begin{figure}
{\epsfxsize\hsize \epsfbox{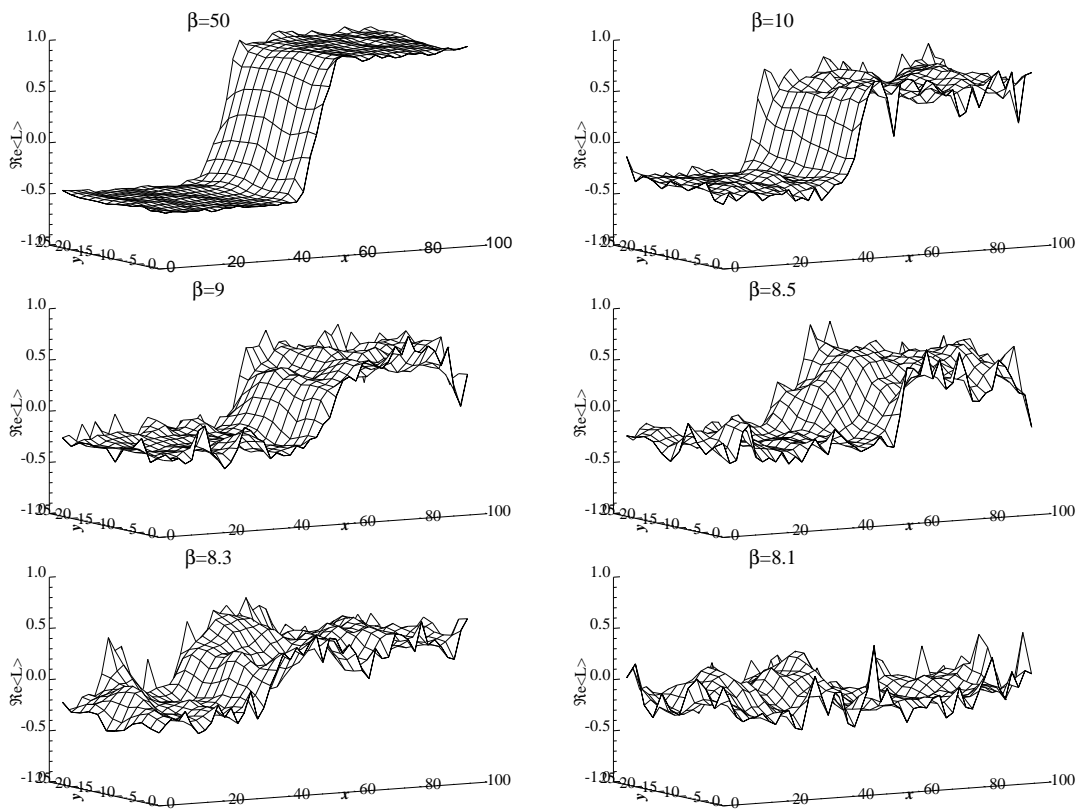}}
\caption{Smoothed Polyakov line profiles, produced by processing the
raw pictures of fig.2. As in that figure, $\beta=50, 10, 9,
8.5, 8.3$ and $8.1$ respectively.}
\end{figure}

However, there is a drawback with this smoothing technique
 as it averages out a whole class of
fluctuations which may be important in the interface collapse.
A more satisfactory method is to produce a contour map from the real
part of the Polyakov lines on
the lattice. Consider following a contour whose
height is mid-way between that of the average Polyakov line at the low-$x$
end of the lattice and that at the high-$x$ end. This  picks out
the mid-height of the interface as it goes across the lattice, as well
as any bubbles of fluctuating phase which are large enough to cross
the contour height. The crucial point to note is that {\it only} the $Z(3)$
interface will cross the entire lattice, {\it i.e.} the only contour
that will wrap once around the lattice in the transverse ($y$)
direction is that 
corresponding to the interface. Any contour representing a bubble of phase will join up with
itself without a net crossing of any lattice boundary. This enables us  to tell the interface apart from any
fluctuations, and to identify the position of its mid-height. By
following contours at various heights between the two extremes, we can
study more of the structure of the interface. The advantage of using contours
is that no prior processing of the Polyakov line data is
required. We can also study contours of the box-car-averaged data
and this can be useful very close to the
critical temperature.

\begin{figure}
{\epsfxsize\hsgraph \epsfbox{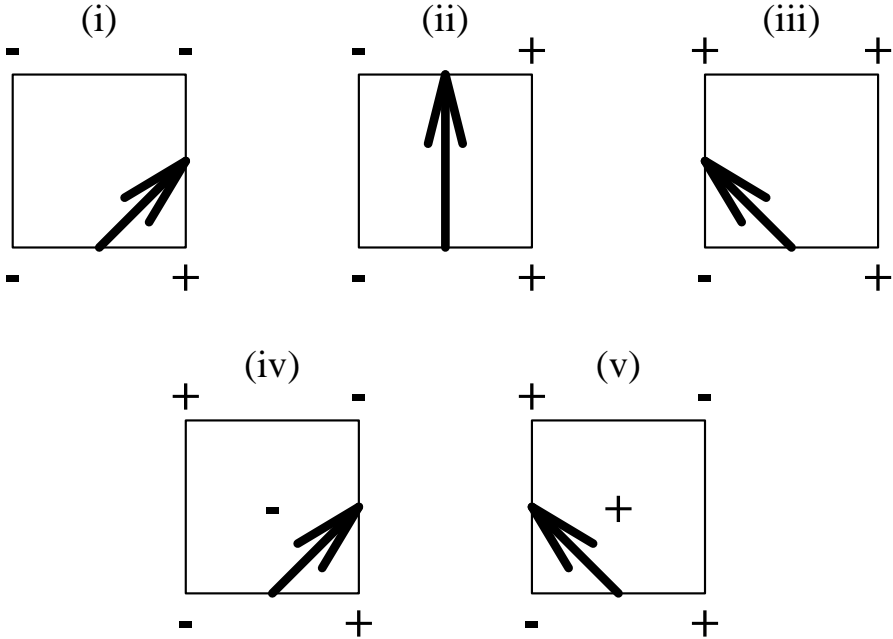}}
\caption{Options for the contour-following algorithm. Given a point
at the desired contour level on the lower side, the Polyakov lines at
the upper two corners determine the progress of the contour in cases
(i)-(iii). A plus (minus) sign indicates a Polyakov line value greater
(less) than
the contour level. The same corner configurations of
(iv) and (v) give two
possible contour directions; we choose between them by considering the
central average of the corner values. These pictures exhaust all
possibilities for the values of the lower corners given, but any general
configuration can be rotated and reflected into one of these.}
\end{figure}

To find a contour at a particular level, we return to the raw Polyakov
line data. For $y=0$, the algorithm follows a procedure similar to that above:
starting from the centre, $x=L_x/2$, it looks for the nearest point passing
through the contour level. Then, it considers the square of
points formed by the two neighbouring lattice sites at $y=0$ and the
corresponding sites at $y=1$.
The algorithm decides to which side of the square the contour should
go, as detailed in fig.4. Then,
starting with this side, the whole procedure continues. Eventually,
the contour must join up with itself. Repeated application of this technique
for different contour levels determines the structure of the interface
to the desired precision. The same procedure can be applied to the
smoothed Polyakov line configurations, to obtain smoothed contour
maps.
\begin{figure}
{\epsfxsize\hsize \epsfbox{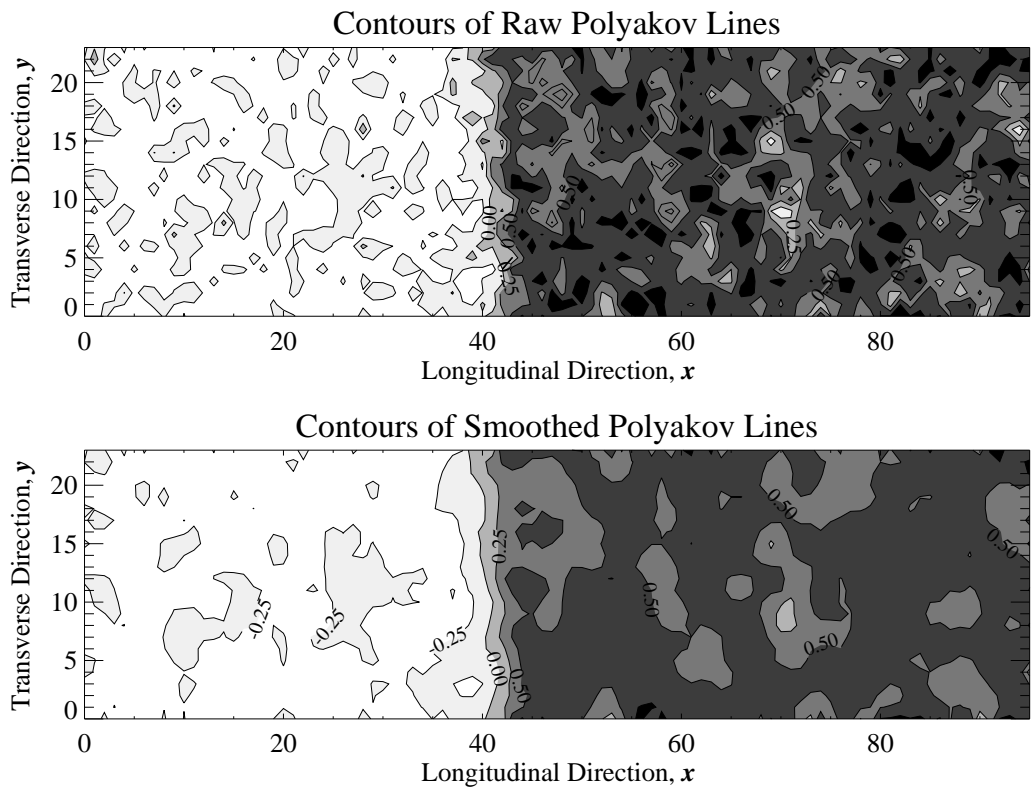}}
\caption{Contour maps of raw and processed Polyakov lines for
$\beta=10$ on a $2\times 24\times 96$ lattice. These maps correspond
to the $\beta=10$ pictures of fig.2 and fig.3 respectively. The
contour levels are at $-0.25, 0, 0.25, 0.5$ and $0.75$, darker regions
corresponding to higher values.}
\end{figure}
An example of a contour map is shown in fig.5, for the same $\beta=10$ gauge
configuration featured in fig. 2 and fig. 3. Note that the interface
in the centre of the maps is represented by
the only contours going all the way across the lattice. The first
picture shows the raw data of fig.2; the second, the smoothed data of
fig.3.

 Left to its own devices, the interface will
execute a random walk in the $x$ direction. When the interface reaches
$x\sim0$ or $x\sim L_x$ it will interact
with the twist and  tunnel into
different $Z(3)$ vacua, producing a different $Z(3)$ interface each
time. However
to study the profile of one particular $Z(3)$ interface without
interference from other vacua the interface should be held in place well away from the
twist, so that no part of it, even its wildest fluctuation, passes
through the twist.  After each Monte-Carlo (MC) sweep of the lattice, we
locate the position of the interface,
and then slide the whole gauge configuration along the
lattice until the interface is re-centred at $x\approx L_x/2$. Any
link variables which
are shifted through the twist are multiplied by the appropriate
factor, so there is no physical
effect and no violation of translation invariance arising from this procedure.

For our MC simulations, we used a mix of Cabibbo-Marinari heat-bath
steps \cite{IIRiv}, using the Kennedy-Pendleton
algorithm \cite{IIRii} to update $SU(2)$
subgroups, and Brown-Woch over-relaxation steps \cite{IIRvi}, after using
many initial heat-bath sweeps to equilibrate the system.
Measurements of
physical operators took place only every four sweeps, to reduce the
correlations caused by adjacent configurations. Since we were
interested in the evolution of the interface,
contour measurements were taken after every sweep of the
lattice. In order to study finite-length effects for the interface, we
performed simulations on a number of different lattice widths
($L_y=18,24,30,36,42,48,54$) and length $L_x=72$,
chosen after studies with various lengths
to ensure that not even the wildest fluctuations of the interface would
reach as far as the twist during our full-length runs. We used $\beta$ values
of 8.25, 8.35, 8.50, 8.75 and 9.00 to study the behaviour of the
interface close to $\beta_c$. The simulations
consisted, in each case, of 2k heat-bath sweeps and 100k main
sweeps. Approximately 100---200 hours of CPU time on a DEC 2100 A500MP
machine were needed for each value of $\beta$.

The statistical analysis of our data, and especially of
 the Wick-subtracted moments which are not expected to be 
normally distributed, is quite complicated and requires a method 
 which does not rely on {\it a priori} knowledge of
 their distribution. We used methods based on the  bootstrap principle 
\cite{RIVi} and full details are given in  \cite{RNi}.

\section{The Fluctuation Moments of the Interface}

\begin{figure}
{\epsfxsize\hsize \epsfysize\vsgraph \epsfbox{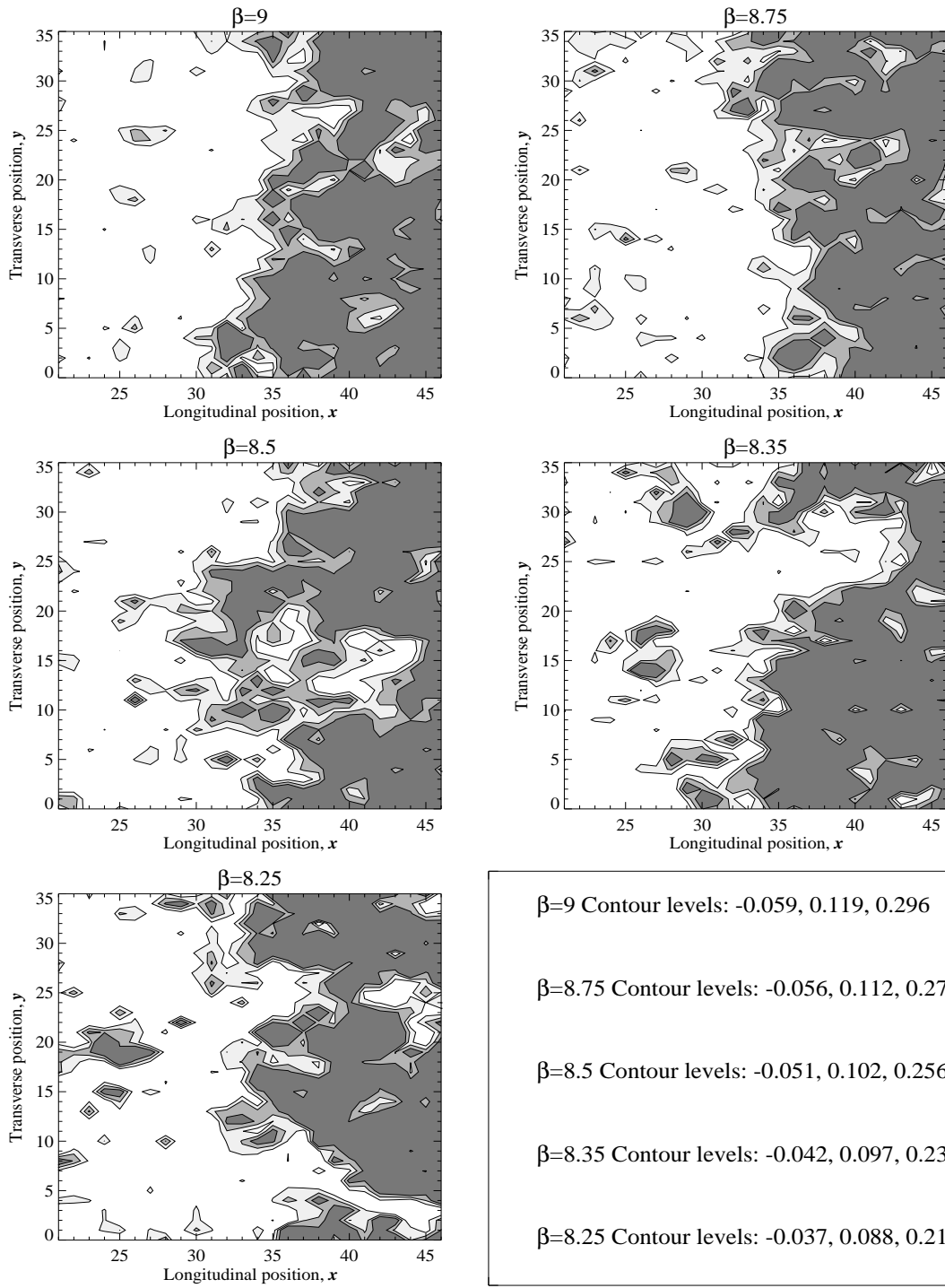}}
\caption{Contour maps of the interface on a $2\times 36\times 72$
lattice, for decreasing temperature. Moving left to right, and then
top to bottom, the picture illustrates the interface structure around the
longitudinal centre of the lattice after
2k heat-bath and 60k main sweeps for $\beta=9, 8.75, 8.5, 8.35$ and $8.25$.}
\end{figure}

In fig.6, a contour snapshot of the interface is shown for each
temperature studied, at the same point in each simulation, just over
halfway through. Three contours have been followed in each
case. We can see that the contours stay close together across most of
the interface, indicating that the interface remains relatively narrow
as the temperature drops. The
shape of the interface appears to fluctuate rather more wildly with the
falling temperature, but the interface does not
spread out noticeably in width.

As discussed in  section 2, the fluctuations are characterised by their
moments $<\phi(y)^n>_c$; $\phi(y)$ can be the displacement from its mean
position of {\it any} contour that has been identified.
We measured the first six moments,
and took separate sets of data  for each of
the three contours followed, at 25\%, 50\% and 75\% of the height of the
interface respectively. The qualitative behaviour is roughly the same
for each contour. The upper and lower contours are subject to
rather more fluctuation than the middle one, as we expect since they
are closer to the outside of the interface and therefore more susceptible
to bubbles of phase forming nearby and distorting the edge of the
interface. The even moments behave very smoothly, whilst the odd moments are very much smaller
and show more variation;  the first moment must average to zero by definition,
 and we  expect all odd
moments to tend to zero for infinite statistics.
We also computed the behaviour of the moments of smoothed data and
found that it is very similar to
that of the raw data, both qualitatively and
quantitatively. This indicates that the smoothing procedure does not
lose a significant amount of information about the fluctuations and is
consistent with the notion that the important fluctuations take place
on a distance scale much larger than the lattice spacing.

Our results for $\beta\geq 8.35$ show reasonable statistical distributions
for the moments over the 100k sweeps. However, those for $\beta=8.25$
and smaller $L_y$ occasionally contain ``spikes'', observations for the
moments which are far larger than we would expect, sometimes by
several magnitudes.
The spikes are not caused by
wild fluctuations passing through the twist, since similar spikes are
not seen for the widest lattices where
fluctuations are larger. They are an artefact of the greater energetic instability
of the smaller interfaces near the critical point.  If parts of the
interface collapse during interaction with nearby bubbles of phase,
the interface contour may briefly undergo a drastic distortion or
partial collapse, even appearing to
pass through the twist because of the change in
Polyakov values there and confusing our algorithm.
We cannot prevent the spikes occurring, since the cause is a real 
physical process, but
they can have a serious effect on the overall averages,
especially for the higher moments. Thus, we would like to be able to
examine the data with the spikes removed. To do this, we impose an
upper cut-off on the statistical distributions.
To ensure that only the spikes are removed, we
need to define the cut-off in a statistically sensible manner. We
take it, for each distribution, to be the median plus 2.5 times the
interquartile range of the distribution of {\it logarithmic} values of
data, where the interquartile range is defined to run between the 25th and 75th
percentiles of the distribution. This ensures that the cut-off
corresponds only to the
upper tip (to be precise, the top 0.04\%) of a normal distribution,
ensuring that only data sets containing spikes are affected by the procedure.
The exact multiplier ``2.5'' can, of course, be varied, but we find
that a lower number ({\it e.g.} 2.0, cutting off 0.3\% of the
distribution), has too great an effect on the normal contribution to
the distributions, whereas a higher number ({\it e.g.} 3.0, cutting
off only 0.002\% of the normal distribution) permits too much
influence from the spikes. The procedure is fully detailed in \cite{RNi}.
In our tabulated results values which were obtained with this procedure
in effect are marked with a \dag.

\medskip\boxit{
\vbox{\tabskip=0pt \offinterlineskip
\def\tabletitles#1#2#3#4#5#6#7#8#9{\DefWarn\BTi\xdef\BTi{table 1:~~
}&&\multispan{17}\hfil {Table 1: Estimates of $<\phi^2>$ from Raw Data (Corrected), 50\% Contour}\hfil&\cr
\tableedge\tablerule\tablegap\tablerule\tablelines
&&#1&&#2&&#3&&#4&&#5&&#6&&#7&&#8&&#9&\cr}
\def\tablerule{\noalign{\hrule}}
\def\tablelines{\omit&height2pt&\omit&height2pt&\omit&height2pt&\omit&height2pt&\omit&height2pt&\omit&height2pt&\omit&height2pt&\omit&height2pt&\omit&height2pt&\omit&height2pt\cr}
\def\tableedge{\omit&height2pt&\multispan{17}&\cr}
\def\tablegap{\omit&\omit\vbox to 3pt{}&\multispan{17}\cr}
\halign to\hsbox{\strut#& \vrule#\tabskip=0em plus 2em& \hfil#\hfil &
\vrule# & \hfil#\hfil & \vrule# & \hfil#\hfil & \vrule# & \hfil#\hfil &
\vrule# & \hfil#\hfil & \vrule# & \hfil#\hfil & \vrule# & \hfil#\hfil &
\vrule# & \hfil#\hfil & \vrule# & \hfil#\hfil &
\vrule#\tabskip=0pt\cr\tablerule\tableedge
\tabletitles{$\beta\ddots L_y$}{\it 18}{\it 24}{\it 30}{\it
36}{\it 42}{\it 48}{\it 54}{\it Slope}
\tablelines\tablerule\tablelines
&&{\it 8.25}&&$5.44^{+0.05}_{-0.04}$&&$8.09^{+0.06}_{-0.06}$&&$11.9^{+0.1}_{-0.1}$&&$14.3^{+0.1}_{-0.1}$&&$17.8^{+0.1}_{-0.1}$&&$20.8^{+0.2}_{-0.2}$&&$23.1^{+0.2}_{-0.2}$&&${}_{ 1.33\pm 0.02}$&\cr
\tablelines\tablerule\tablelines
&&{\it 8.35}&&$4.52^{+0.03}_{-0.03}$&&$6.69^{+0.05}_{-0.05}$&&$9.13^{+0.07}_{-0.06}$&&$11.4^{+0.1}_{-0.1}$&&$13.4^{+0.1}_{-0.1}$&&$16.5^{+0.1}_{-0.1}$&&$18.2^{+0.1}_{-0.1}$&&${}_{ 1.27\pm 0.01}$&\cr
\tablelines\tablerule\tablelines
&&{\it 8.50}&&$3.93^{+0.03}_{-0.03}$&&$5.19^{+0.03}_{-0.03}$&&$7.17^{+0.05}_{-0.05}$&&$8.89^{+0.07}_{-0.06}$&&$10.0^{+0.1}_{-0.1}$&&$11.9^{+0.1}_{-0.1}$&&$13.8^{+0.1}_{-0.1}$&&${}_{ 1.15\pm 0.01}$&\cr
\tablelines\tablerule\tablelines
&&{\it 8.75}&&$3.06^{+0.02}_{-0.02}$&&$4.21^{+0.03}_{-0.03}$&&$5.32^{+0.04}_{-0.03}$&&$6.34^{+0.04}_{-0.04}$&&$7.62^{+0.05}_{-0.05}$&&$8.85^{+0.06}_{-0.06}$&&$9.57^{+0.06}_{-0.06}$&&${}_{ 1.05\pm 0.01}$&\cr
\tablelines\tablerule\tablelines
&&{\it 9.00}&&$2.62^{+0.02}_{-0.02}$&&$3.44^{+0.02}_{-0.02}$&&$4.38^{+0.03}_{-0.03}$&&$5.27^{+0.03}_{-0.03}$&&$6.29^{+0.04}_{-0.04}$&&$7.37^{+0.05}_{-0.05}$&&$7.89^{+0.05}_{-0.05}$&&${}_{ 1.03\pm 0.01}$&\cr
\tablelines\tablerule\tablelines
&&{\it Slope}&&${}_{-0.31\pm 0.03}$&&${}_{-0.36\pm 0.03}$&&${}_{-0.42\pm 0.03}$&&${}_{-0.42\pm 0.03}$&&${}_{-0.44\pm 0.02}$&&${}_{-0.45\pm 0.03}$&&${}_{-0.46\pm 0.04}$&&&\cr
\tablelines\tablerule}}}\bigskip

\bigskip\boxit{
\vbox{\tabskip=0pt \offinterlineskip
\def\tabletitles#1#2#3#4#5#6#7#8#9{\DefWarn\BTiv\xdef\BTiv{table 2:
}&&\multispan{17}\hfil {Table 2:
 Estimates of $<\phi^4>_{C}$ from Raw Data (Corrected), 50\% Contour}\hfil&\cr
\tableedge\tablerule\tablegap\tablerule\tablelines
&&#1&&#2&&#3&&#4&&#5&&#6&&#7&&#8&&#9&\cr}
\def\tablerule{\noalign{\hrule}}
\def\tablelines{\omit&height2pt&\omit&height2pt&\omit&height2pt&\omit&height2pt&\omit&height2pt&\omit&height2pt&\omit&height2pt&\omit&height2pt&\omit&height2pt&\omit&height2pt\cr}
\def\tableedge{\omit&height2pt&\multispan{17}&\cr}
\def\tablegap{\omit&\omit\vbox to 3pt{}&\multispan{17}\cr}
\halign to\hsbox{\strut#& \vrule#\tabskip=0em plus 2em& \hfil#\hfil &
\vrule# & \hfil#\hfil & \vrule# & \hfil#\hfil & \vrule# & \hfil#\hfil &
\vrule# & \hfil#\hfil & \vrule# & \hfil#\hfil & \vrule# & \hfil#\hfil &
\vrule# & \hfil#\hfil & \vrule# & \hfil#\hfil &
\vrule#\tabskip=0pt\cr\tablerule\tableedge
\tabletitles{$\beta\ddots L_y$}{\it 18}{\it 24}{\it 30}{\it
36}{\it 42}{\it 48}{\it 54}{\it Slope}
\tablelines\tablerule\tablelines
&&{\it 8.25}&&$63.2^{+3.7}_{-3.0}$&&$115^{+7}_{-5}$&&$273^{+14}_{-11}$&&$326^{+9}_{-9}$&&$425^{+15}_{-14}$&&$580^{+14}_{-13}$&&$658^{+15}_{-15}$&&${}_{ 2.12\pm 0.07}$&\cr
\tablelines\tablerule\tablelines
&&{\it 8.35}&&$30.6^{+1.0}_{-1.1}$&&$61.7^{+2.1}_{-2.0}$&&$108^{+3}_{-3}$&&$166^{+6}_{-5}$&&$197^{+6}_{-6}$&&$322^{+1}_{-1}$&&$299^{+10}_{-9}$&&${}_{ 2.17\pm 0.06}$&\cr
\tablelines\tablerule\tablelines
&&{\it 8.50}&&$19.2^{+0.6}_{-0.5}$&&$27.8^{+0.8}_{-0.8}$&&$59.0^{+2.0}_{-1.8}$&&$82.0^{+2.8}_{-2.5}$&&$87.8^{+2.8}_{-2.8}$&&$126^{+5}_{-4}$&&$163^{+6}_{-5}$&&${}_{ 1.97\pm 0.05}$&\cr
\tablelines\tablerule\tablelines
&&{\it 8.75}&&$9.68^{+0.34}_{-0.31}$&&$16.3^{+0.6}_{-0.6}$&&$22.3^{+0.7}_{-0.6}$&&$30.3^{+1.1}_{-1.0}$&&$43.0^{+2.2}_{-2.0}$&&$48.2^{+1.9}_{-1.7}$&&$49.7^{+1.9}_{-1.8}$&&${}_{ 1.55\pm 0.04}$&\cr
\tablelines\tablerule\tablelines
&&{\it 9.00}&&$5.84^{+0.17}_{-0.16}$&&$8.76^{+0.25}_{-0.25}$&&$13.0^{+0.4}_{-0.4}$&&$15.6^{+0.6}_{-0.5}$&&$23.3^{+0.8}_{-0.8}$&&$28.6^{+1.1}_{-1.0}$&&$28.5^{+1.3}_{-1.0}$&&${}_{ 1.54\pm 0.03}$&\cr
\tablelines\tablerule\tablelines
&&{\it Slope}&&${}_{-0.99\pm 0.06}$&&${}_{-1.10\pm 0.07}$&&${}_{-1.28\pm 0.07}$&&${}_{-1.25\pm 0.11}$&&${}_{-1.21\pm 0.06}$&&${}_{-1.27\pm 0.11}$&&${}_{-1.28\pm 0.12}$&&&\cr
\tablelines\tablerule}}}\bigskip

\bigskip\boxit{
\vbox{\tabskip=0pt \offinterlineskip
\def\tabletitles#1#2#3#4#5#6#7#8#9{\DefWarn\BTvii\xdef\BTvii{table 3:
}&&\multispan{17}\hfil {Table 3:
 Estimates of $<\phi^6>_{C}$ from Raw Data (Corrected), 50\% Contour}\hfil&\cr
\tableedge\tablerule\tablegap\tablerule\tablelines
&&#1&&#2&&#3&&#4&&#5&&#6&&#7&&#8&&#9&\cr}
\def\tablerule{\noalign{\hrule}}
\def\tablelines{\omit&height2pt&\omit&height2pt&\omit&height2pt&\omit&height2pt&\omit&height2pt&\omit&height2pt&\omit&height2pt&\omit&height2pt&\omit&height2pt&\omit&height2pt\cr}
\def\tableedge{\omit&height2pt&\multispan{17}&\cr}
\def\tablegap{\omit&\omit\vbox to 3pt{}&\multispan{17}\cr}
\halign to\hsbox{\strut#& \vrule#\tabskip=0em plus 2em& \hfil#\hfil &
\vrule# & \hfil#\hfil & \vrule# & \hfil#\hfil & \vrule# & \hfil#\hfil &
\vrule# & \hfil#\hfil & \vrule# & \hfil#\hfil & \vrule# & \hfil#\hfil &
\vrule# & \hfil#\hfil & \vrule# & \hfil#\hfil &
\vrule#\tabskip=0pt\cr\tablerule\tableedge
\tabletitles{$\beta\ddots L_y$}{\it 18}{\it 24}{\it 30}{\it
36}{\it 42}{\it 48}{\it 54}{\it Slope}
\tablelines\tablerule\tablelines
&&{\it 8.25}&&$3530$&&$6330$&&$23500$&&$18500$&&$18500$&&$14500$&&$15200$&&${}_{ 1.46\pm 0.22}$&\cr
&& &&${}^{+490}_{-400}$&&${}^{+1080}_{-990}$&&${}^{+3500}_{-2800}$&&${}^{+1600}_{-1500}$&&${}^{+3400}_{-2600}$&&${}^{+2800}_{-2300}$&&${}^{+3400}_{-2900}$&& &\cr
\tablelines\tablerule\tablelines
&&{\it 8.35}&&$712$&&$1700$&&$2860$&&$5180$&&$4790$&&$10200$&&$2640$&&${}_{ 2.38\pm 0.29}$&\cr
&& &&${}^{+75}_{-55}$&&${}^{+210}_{-190}$&&${}^{+440}_{-390}$&&${}^{+670}_{-580}$&&${}^{+1050}_{-840}$&&${}^{+2100}_{-1700}$&&${}^{+2300}_{-1680}$&& &\cr
\tablelines\tablerule\tablelines
&&{\it 8.50}&&$259^{+29}_{-23}$&&$350^{+85}_{-73}$&&$1090^{+160}_{-150}$&&$1710^{+350}_{-280}$&&$912^{+316}_{-250}$&&$1580^{+430}_{-360}$&&$1020^{+670}_{-510}$&&${}_{ 1.84\pm 0.22}$&\cr
\tablelines\tablerule\tablelines
&&{\it 8.75}&&$82.5^{+9.9}_{-8.6}$&&$142^{+40}_{-35}$&&$102^{+30}_{-24}$&&$61.5^{+46.0}_{-43.4}$&&$244^{+116}_{-102}$&&$17.3^{+101.8}_{-78.0}$&&$-634^{+111}_{-91}$&&${}_{?\pm?}$&\cr
\tablelines\tablerule\tablelines
&&{\it 9.00}&&$24.7^{+3.6}_{-3.1}$&&$34.4^{+9.1}_{-9.5}$&&$27.3^{+15.8}_{-13.9}$&&$-35.7^{+19.5}_{-16.2}$&&$-65.2^{+34.2}_{-28.0}$&&$-214^{+47}_{-39}$&&$-139^{+61}_{-47}$&&${}_{ 0.49\pm 0.11}$&\cr
\tablelines\tablerule\tablelines
&&{\it Slope}&&${}_{-1.98\pm 0.11}$&&${}_{-2.10\pm 0.15}$&&${}_{-2.58\pm 0.25}$&&${}_{-1.94\pm 0.60}$&&${}_{-2.09\pm 0.12}$&&${}_{-2.79\pm 0.80}$&&${}_{-1.87\pm 0.07}$&&&\cr
\tablelines\tablerule}}}\eject

\begin{figure}
{\epsfxsize\hsize \epsfbox{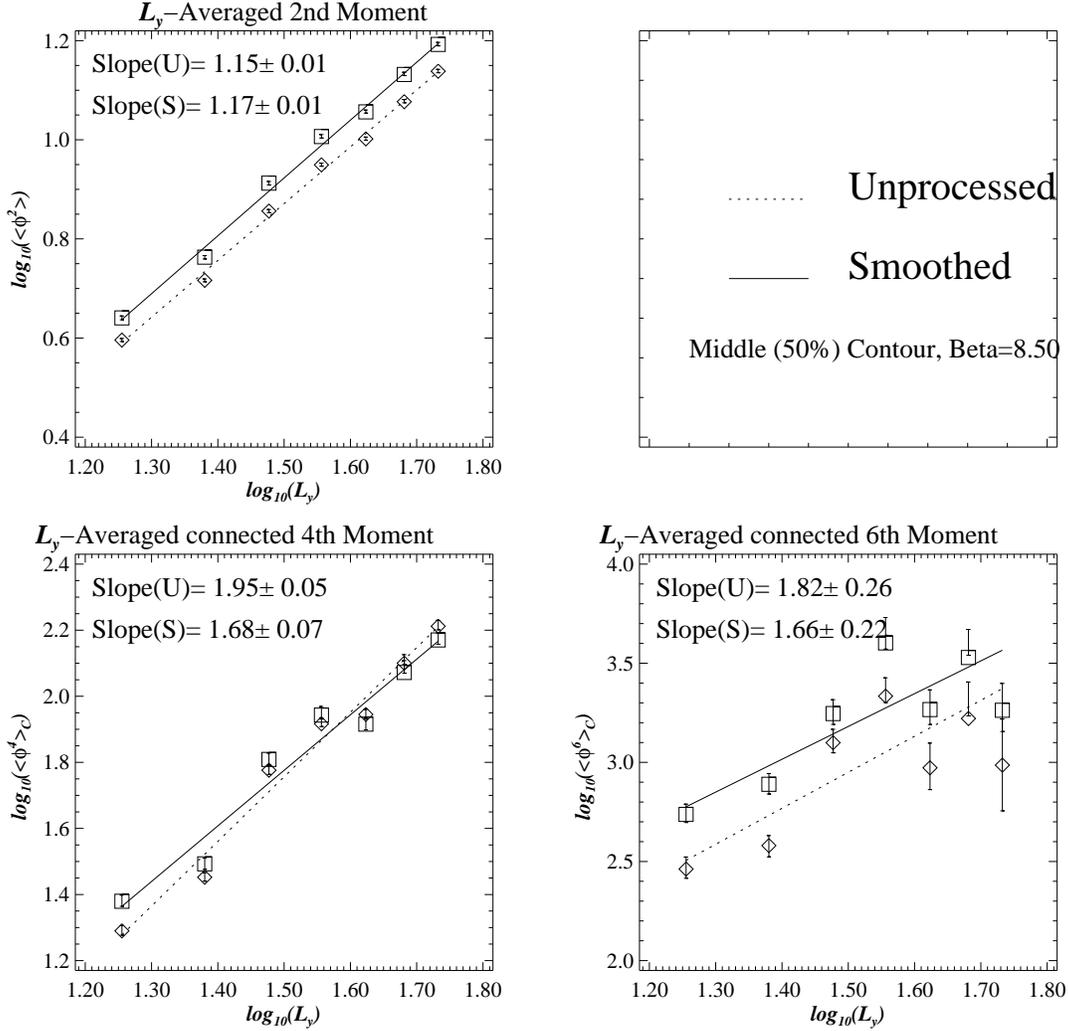}}
\caption{The connected moments  at $\beta=8.5$ plotted against
the lattice width. Diamonds and boxes represent raw and smoothed data
respectively. The lines are best fits to (8).}
\end{figure}

Fig.7 shows the dependence of the moments, calculated for both the raw and
 the smoothed 50\% contour, on the lattice width at $\beta=8.5$. The slopes
for the second and fourth moments are in reasonable agreement with the 
simple model results \rf{moments} although
 the agreement
is not good for the
sixth.
\begin{figure}
{\epsfxsize\hsize \epsfbox{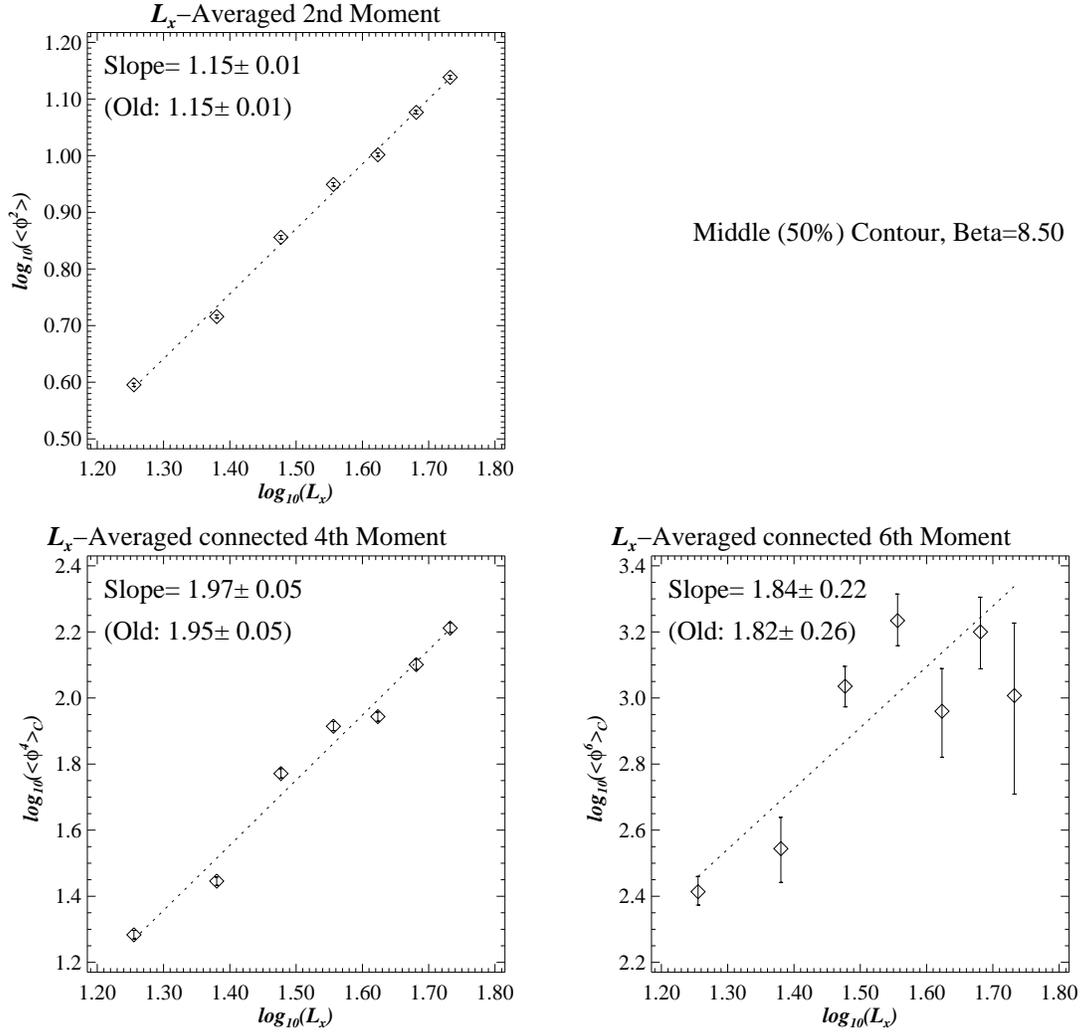}}
\caption{The connected moments with spike suppression at $\beta=8.5$.}
\end{figure}
 In fig.8 we show the same data but with spike suppression in operation.
The results are very similar to those shown in fig.7 demonstrating that
the spikes are not a significant problem at $\beta=8.5$.
The roughness exponents defined by
\be <\phi^n>_c\sim L_y^{n\alpha_n}\label{hohum}\ee
are tabulated in table 4 for all the $\beta$ values we have studied
(an asterisk means that no statistically significant determination
was possible).

\begin{table}
\begin{center}
Table 4
\end{center}
\begin{center}
\begin{tabular}{|c|c|c|c|}
\hline
$\beta$&$2\alpha_2$&$4\alpha_4$&$6\alpha_6$\\
\hline
8.25&1.33(2)\dag&2.12(7)\dag&1.46(22)\dag\\
\hline
8.35&1.27(1)\dag&2.17(6)\dag&2.38(29)\dag\\
\hline
8.5&1.15(1)&1.95(5)&1.82(26)\\
\hline
8.75&1.05(1)&1.54(4)&*\\
\hline
9.0&1.03(1)&1.54(3)&*\\
\hline
\end{tabular}
\end{center}
\end{table}

\begin{figure}
{\epsfxsize\hsize \epsfbox{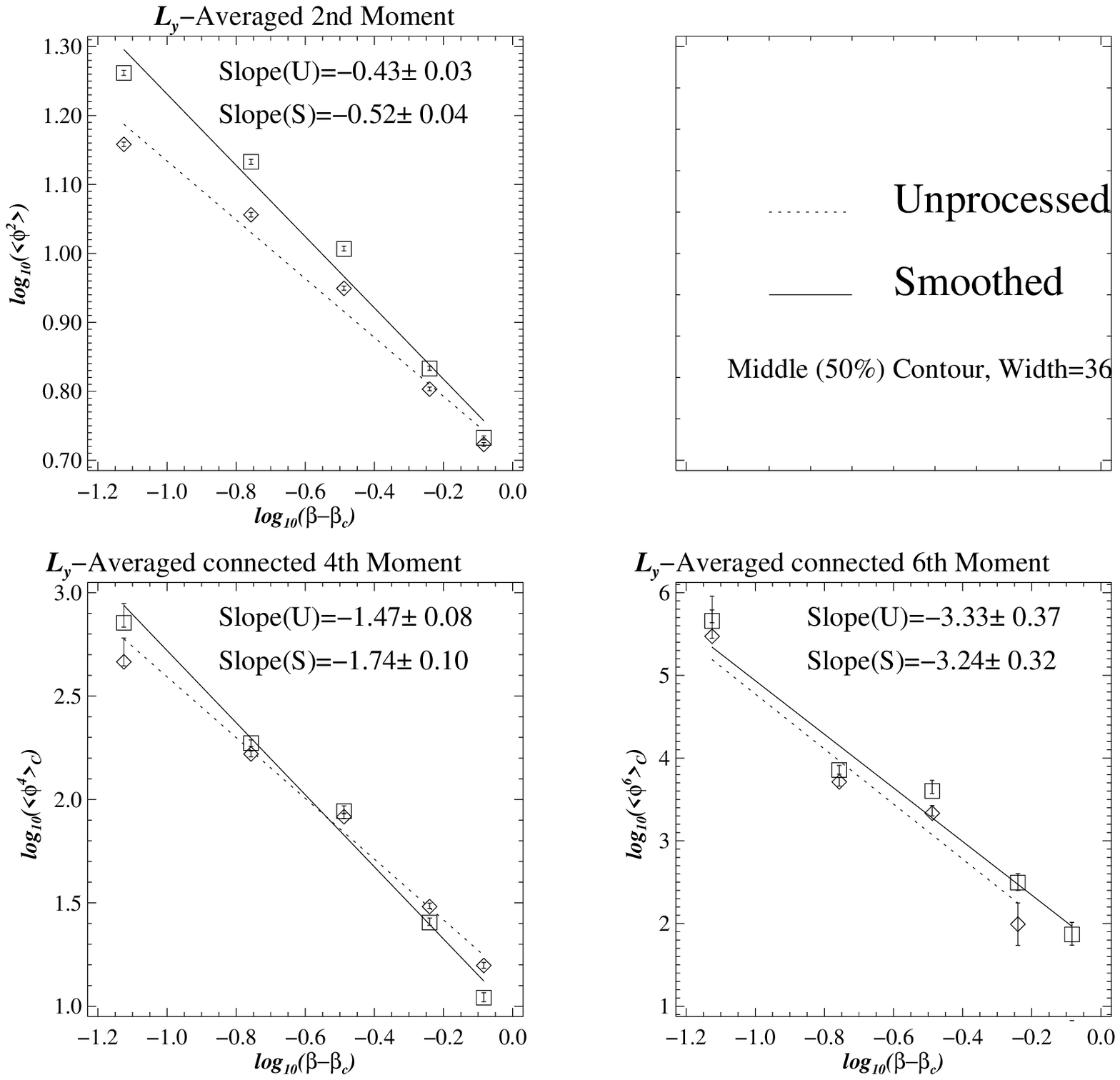}}
\caption{The connected moments  on the $L_y=36$ lattice plotted
against $\beta$. The lines are
 best fits to (9).}
\end{figure}

In fig.9, we plot the moments at fixed $L_y=36$ against $\beta$ when the
spike-suppression procedure discussed above is not operating. The lines are fits
of the form 
\be <\phi^n>_c\sim (\beta-\beta_c)^{-\gamma_n}\label{bscale}\ee 
taking $\beta_c=8.175$.
Comparing the second and fourth moments  with \rf{moments} we find that 
\be\gamma\sim (\beta-\beta_c)^{0.43(3)},\qquad \lambda_0\sim
(\beta-\beta_c)^{0.25(14)}\ee
which then  predict that 
\be<\phi^6>_{c}\sim(\beta-\beta_c)^{-2.51(21)},\ee
 within roughly one
standard deviation of the observed behaviour.
\begin{figure}
{\epsfxsize\hsize \epsfbox{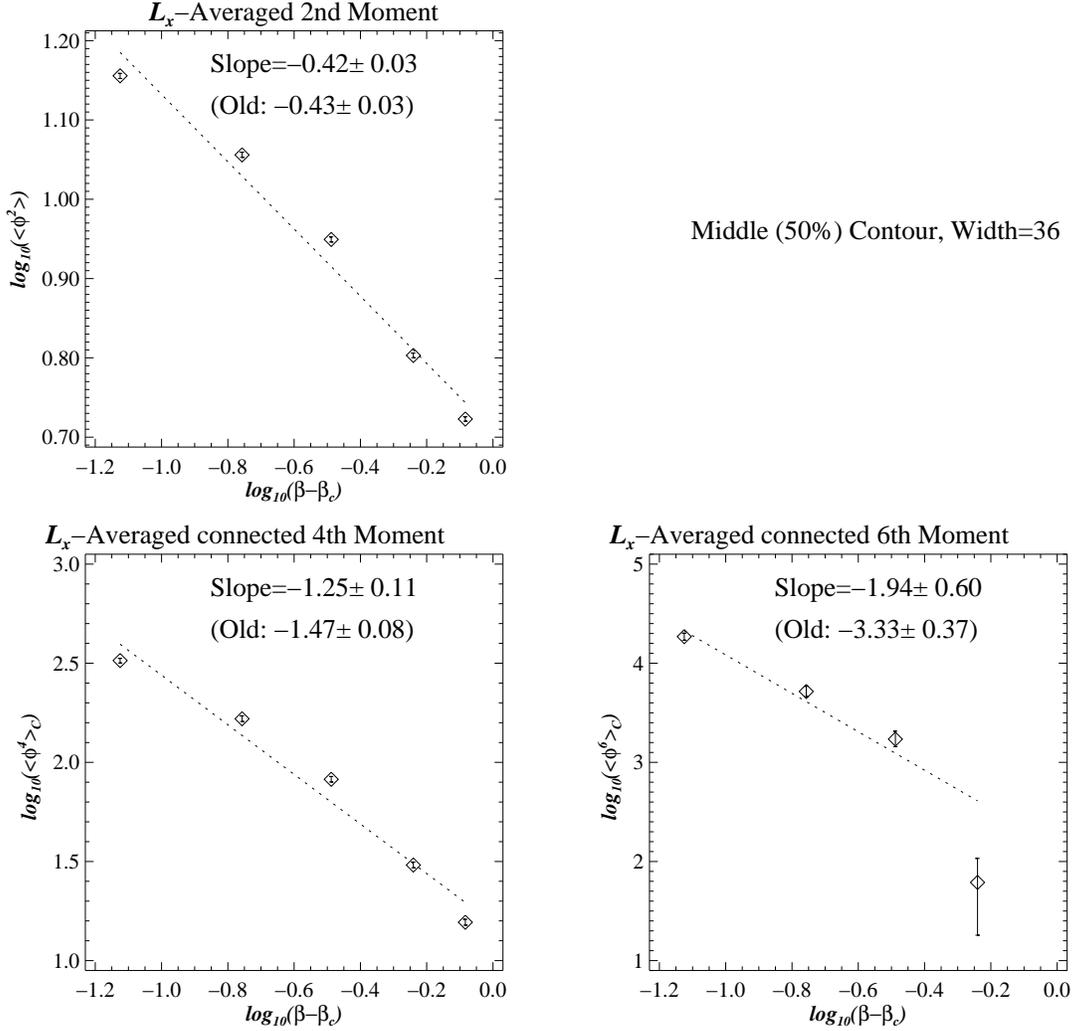}}
\caption{The connected moments with spike suppression  on the 
$L_y=36$ lattice plotted
against $\beta$. The lines are
 best fits to (9).}
\end{figure}
In fig.10  we show the same data as for fig.9 but with spike suppression.
We see that the results for the second moment are scarcely affected 
and the change for the fourth moment is fairly small but the effect
on the sixth moment, which we would expect to be most affected by the 
spikes, is substantial.
\begin{table}
\begin{center}
Table 5
\end{center}
\begin{center}
\begin{tabular}{|c|c|c|c|}
\hline
$L_y$&$\gamma_2$&$\gamma_4$&$\gamma_6$\\
\hline
18&.41(1)&.99(6)\dag&1.98(11)\dag\\
\hline
24&.37(2)&1.10(7)\dag&2.10(15)\dag\\
\hline
30&.44(2)&1.28(7)\dag&2.58(25)\dag\\
\hline
36&.43(3)&1.25(11)\dag&1.96(60)\dag\\
\hline
42&.44(2)&1.29(5)&2.09(12)\dag\\
\hline
48&.45(3)&1.26(12)&2.79(80)\dag\\
\hline
54&.46(4)&1.32(11)&1.87(57)\dag\\
\hline
\end{tabular}
\end{center}
\end{table}
Again comparing the second and fourth moments with \rf{moments}
we find that 
\be\gamma\sim (\beta-\beta_c)^{0.42(3)},\qquad \lambda\sim
(\beta-\beta_c)^{0.43(16)}.\ee
which in turn predict that 
\be<\phi^6>_{C}\sim(\beta-\beta_c)^{-2.08(24)},\ee
 extremely close
to the observed behaviour. The other sets of revised data now give
very consistent results for $\gamma$ and $\lambda$.

 Our other sets of data give
similar results and the $\gamma_n$ are tabulated in table 5 for all
the $L_y$ values we have studied.

\section{The Interface Width, Screening and Mobility}

\begin{figure}
{\epsfxsize\hsgraph \epsfbox{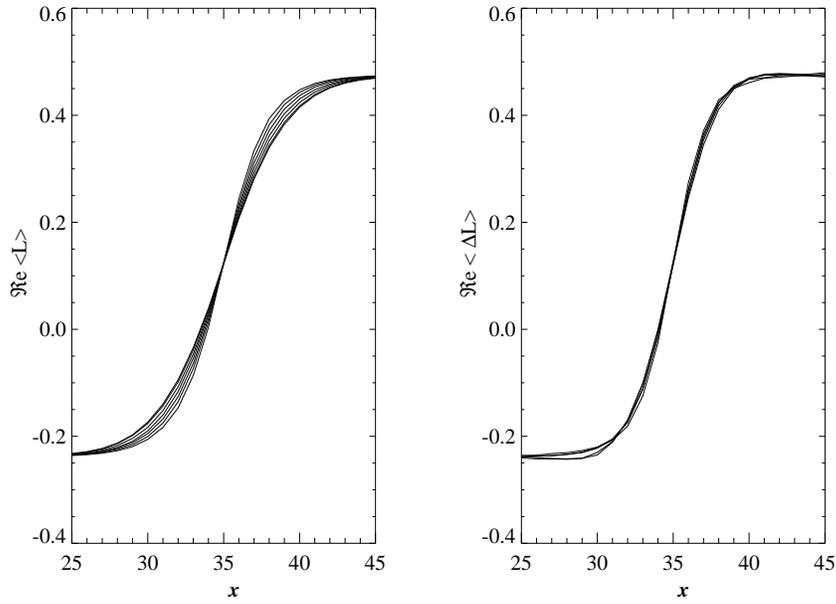}}
\caption{The transverse Polyakov lines at $\beta=9.0$ and the weighted
first differences discussed in the text.}
\end{figure}

We define 
the transverse averaged Polyakov line by
\be\bar L(x,L_y)=L_y^{-1}\sum_y L(x,y)\ee
At very high temperatures where the interface is essentially rigid
we expect it to  show the classic instanton shape and be independent of $L_y$
\cite{IRx}
\be \bar L(x,L_y)=f(x)\sim A\tanh\left({x-x_0\over l}\right)+B\label{good}\ee
where $A$ and $B$ are constants and $l$ is the characteristic width of
the interface.  As the temperature decreases and the interface 
fluctuates more (roughens) some of the spread in $\bar L(x)$ will be
caused by the roughening; provided the fluctuations are not too great,
and assuming that the moments of  $\phi$ are given by the free random 
walk results (good enough at the accuracy of measurement that we have),
we can approximate the measured $\bar L(x,L_y)$ by
\bea\bar L(x,L_y)&=&L_y^{-1}\sum_y<f(x-\phi(y))>_\phi\nn\\
&=&f(x)+L_yf_1(x)+L_y^2f_2(x)+\ldots \eea
\begin{figure}
{\epsfxsize\hsgraph \epsfbox{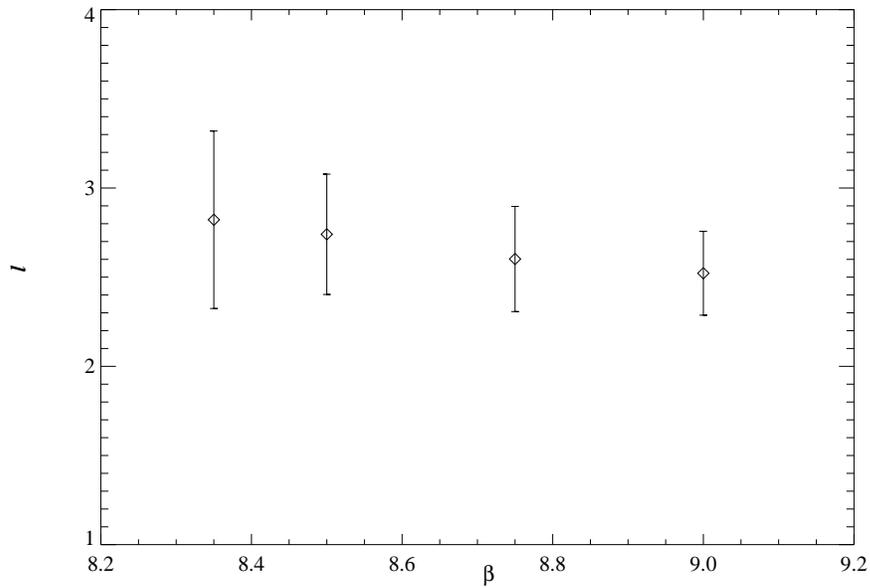}}
\caption{The interface width deduced from transverse Polyakov lines 
plotted against  $\beta$.}
\end{figure}
The piece linear in $L_y$ can be removed by taking weighted first differences
of measurements at different $L_y$,
\bea \Delta \bar L(L_1,L_2)&=&{L_2\bar L(x,L_1)-L_1\bar L(x,L_2)\over L_2-L_1}\nn\\
&=& f(x)+ O(L_{1,2}^2)\eea
 In the  same way both linear and quadratic
$L_y$ dependence can be removed by taking weighted second differences. Fig.11
shows the results of this procedure at $\beta=9.0$; once the part linear
in $L_y$ is removed all the curves collapse quite convincingly  onto one.
Fitting (\ref{good}) to this universal curve a value for $l$ can be extracted.
As $\beta$ is decreased the procedure, which essentially treats the
fluctuations as a perturbation, works less well but the behaviour is 
reasonable down to $\beta=8.35$; it was not possible to measure $l$ at
$\beta=8.25$.  The results for $l$ are plotted in fig.12 and show a mild 
increase as $\beta$ decreases but we do not have enough information
to extract any critical behaviour.

\begin{figure}
{\epsfxsize\hsgraph \epsfbox{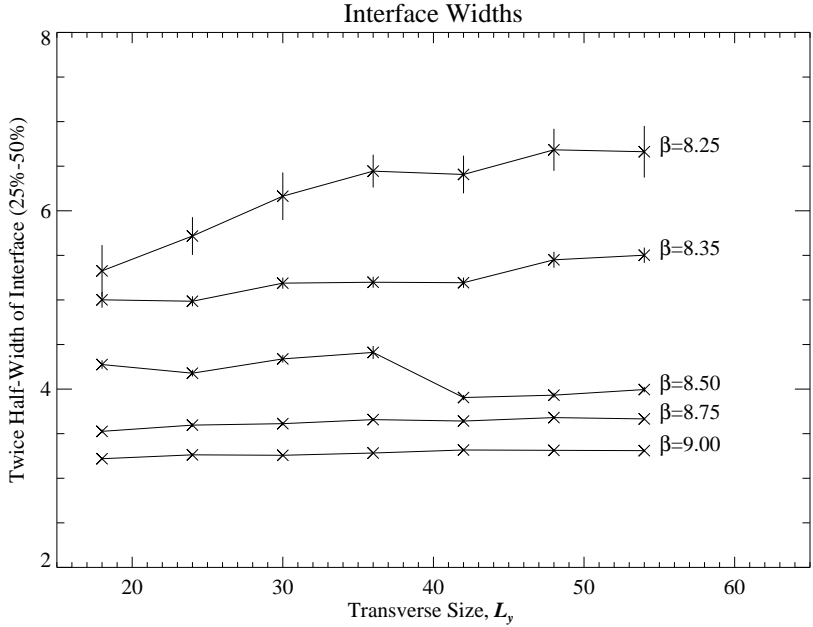}}
\caption{The average
interface width, defined as twice the   average separation of the 25\% and
50\% contours, measured in units of the lattice spacing.}
\end{figure}

We have also measured the interface width by using the
contour data directly. The results of  defining  the width as twice the separation
between 25\% and 50\% are shown in fig.13. Unfortunately we are only able
to measure the separation in the $x$ direction; as the interface fluctuates
this quantity is bound to be greater than the shortest distance between
the contours and so it is inevitable that, especially  at the smaller $\beta$ values,
it will be over-estimated.  Even without taking this into account we see that
the width does not grow rapidly as the temperature decreases towards its critical
value and the two methods for measuring the interface width yield qualitatively
consistent results.

\begin{figure}
{\epsfxsize\hsgraph \epsfbox{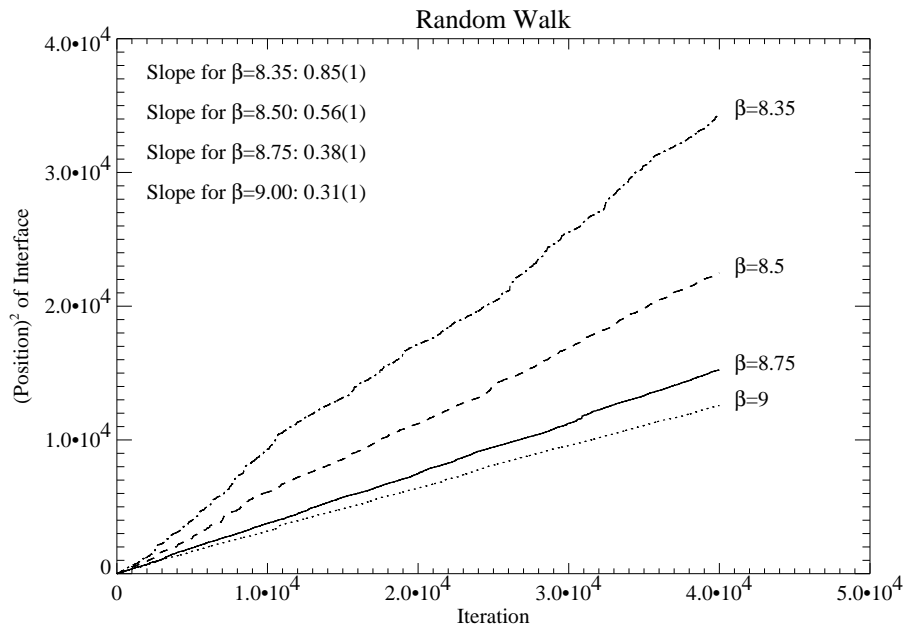}}
\caption{The wandering of the
interface, for temperatures near the critical temperature, over 40k
sweeps after an initial 400 heat-bath sweeps.}
\end{figure}

\begin{figure}
{\epsfxsize\hsgraph \epsfbox{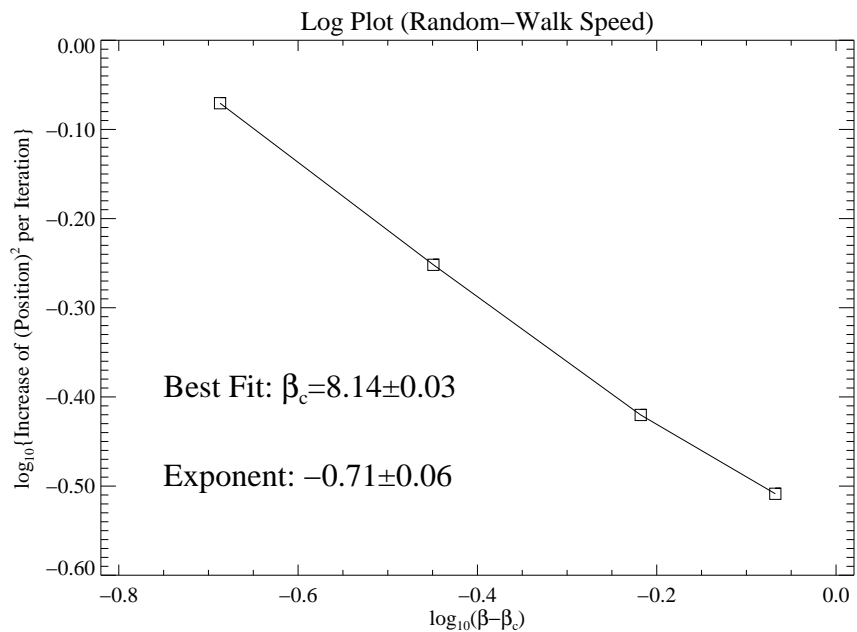}}
\caption{The divergence of the speed of the random walk
in fig.12 as the temperature approaches its  critical value.}
\end{figure}

Finally, as a curiosity, we have examined some quantities
which describe the dynamical (in computer time) behaviour of
the interface and therefore relate to the efficiency of our
MC algorithm in equilibrating the interface itself. 
The speed of the random walk in the $x$-direction made by the
interface can be measured by turning off the mechanism
that constantly re-centres the interface. As the temperature drops and
more and larger bubbles of the ``wrong'' phase form  within the main domains,
the interface will sometimes combine with them and so we might expect it to move
faster.
In fig.14, we do indeed see an increase in speed as the
temperature drops and   the
speed appears to diverge as $\beta\rightarrow\beta_c$.
This is shown more clearly in
fig.15 where we fit the speed to a function of the
form
\be\frac{\del x^2}{\del t} \sim (\beta-\beta_c)^{\beta_{RW}}.\ee
Our results suggest values of
\be \beta_c=8.14(3),\qquad \beta_{RW}=-0.71(6).\ee
This value for $\beta_c$ is  consistent with that obtained by other
methods, and the dynamical exponent governing the critical
acceleration of the random walk is given by $\beta_{RW}$.

Studies of discrete
models \cite{IIIRii} and
continuum growth equations \cite{IIIRiv} in condensed matter systems
make predictions for the behaviour of the width of an interface,
usually defined to be the square-root of our
second moment,
\be W(L_y,t)=\sqrt{\bar{\phi^2}-{\bar\phi}^2},\ee
as a function of the time, $t$, since its formation with initial width
zero.
For times much smaller than some
critical time, $t_X(L_y)$, determined by the interface size, an
exponent $\beta_G$ will govern the growth of the width
\be W(t)\sim t^{\beta_G},\qquad t\ll t_X(L_y).\ee
For times much larger than this  the width is
 governed by $\alpha_2$, the roughness exponent, as we discussed
in section 5.
Fig.16 shows an example of the growth of a
$Z(3)$ interface. At high $t$ the size saturates as usual, but at low
$t$  we see growth governed by an exponent, as in the condensed matter
models. For our interface, the average profile yields
\be\beta_G=0.61(2).\ee

\newpage
%\begin{figure}[p]
{\epsfysize\vsgraph \epsfbox{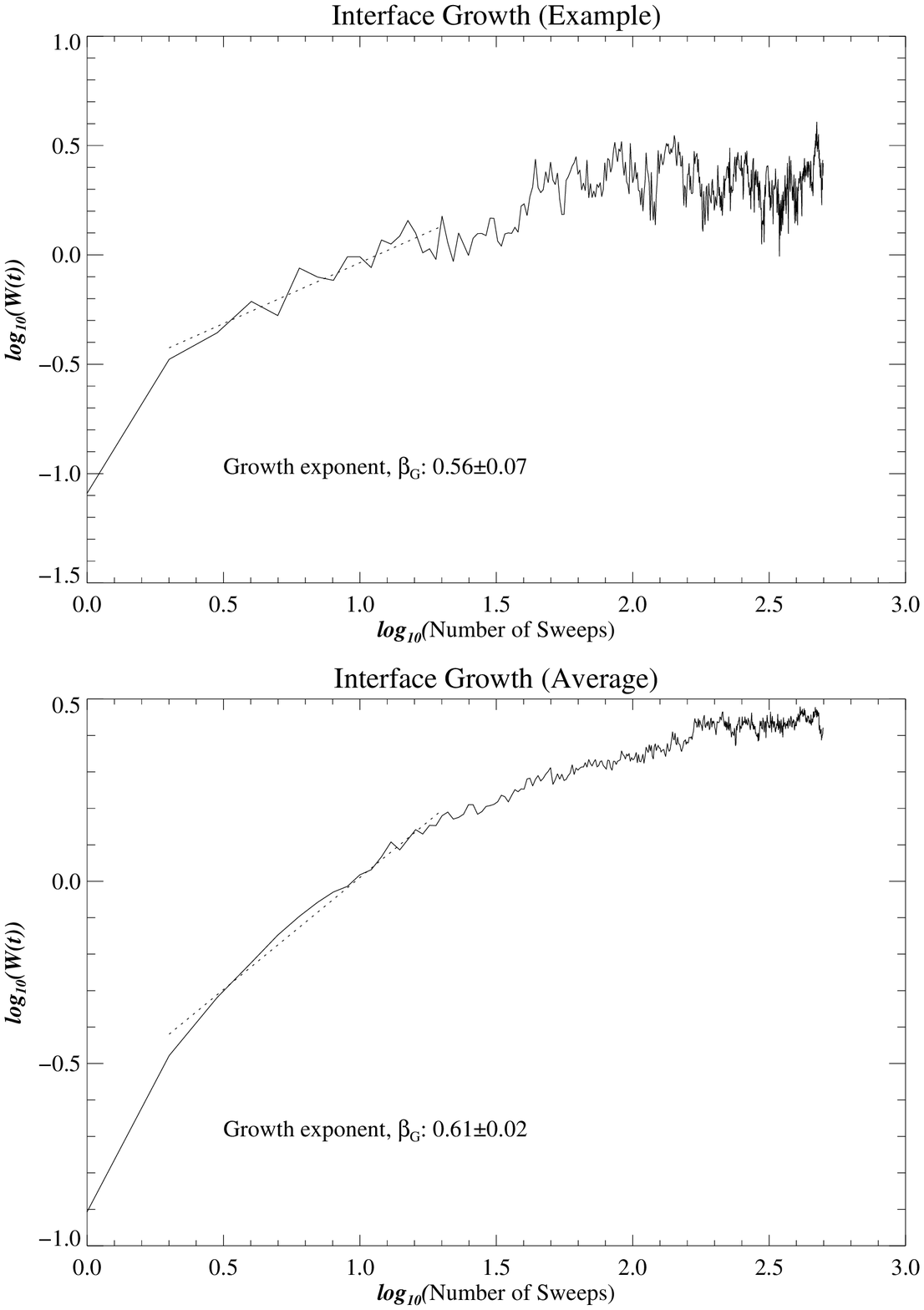}}
\noindent Figure 16:{ An example of the growth in size of an interface for
$\beta=8.5, L_y=36$, and a growth profile averaged over all $L_y$ and
$\beta$. In each case, $\beta_G$ is estimated from the slope of the
dotted line.}
%\end{figure}

\newpage
\section{Conclusions}

Despite the increasing fluctuations and decreasing height of the interface
as the critical temperature is approached we have shown that it is quite 
practicable to isolate it and to measure its geometrical properties. As one
might expect of a second order transition the fluctuation moments diverge
as $\beta\downarrow\beta_c$ in a manner which is to a good approximation independent
of the transverse size of the lattice, $L_y$. That the relation between the
three moments measured is more or less as predicted by the simple model \rf{moments}
may be fortuitous because the model does not describe the dependence of the 
moments on the transverse size very well.  One of the reasons why we are
 able to measure the fluctuations reliably is that the intrinsic width 
remains small; from the measurements we have made there is no real
indication that this quantity does not remain finite as $\beta\downarrow\beta_c$.
The dynamical (in computer time) behaviour of the interface is consistent with the
usual behaviour encountered in other systems.

From this work and \cite{IRxiia,stwjfw} it seems clear that the $Z(N)$ interface
in the Euclidean formulation of the theory really does have all the properties
expected of such an interface in an ordinary statistical mechanical system as
well as being consistent with weak coupling perturbation theory.  The only
objection to such a conclusion is that most of this work has been done at
$L_t=2$.  However the behaviour of the interface free energy in $SU(2)$ at
high temperatures has been studied in \cite{IRxiia} for $L_t=2,3,4$, and $6$
and they find essentially no $L_t$ dependence.

The relation of this picture to Minkowski space is as yet unclear.
In the Euclidean path integral, the different $Z(N)$ vacua, and consequently
domain walls between them, will appear in the ensemble of gauge field 
configurations that is integrated over to yield expectation values. We expect
at least some observable quantities, call them $Q$, to depend upon the presence
of these vacua in the sense that if there were only one vacuum $Q$ would be
different. The observables must be the same when calculated in Minkowski space
as in Euclidean space.  It follows that there must be some 
(necessarily non-perturbative) feature of the 
ensemble of gauge field configurations in Minkowski space which is responsible
for $Q$ taking a different value from the single Euclidean vacuum prediction
(or else there are no $Q$s).  The $Q$ observables potentially correspond 
to interesting physical effects such as baryogenesis and CP violation in the
early universe \cite{CKA,CKB}; these calculations can be criticised on the 
grounds that they assume the Euclidean domain structure carries over to
Minkowski space but it may be that the calculation can be reformulated 
entirely in Euclidean space in which case the results should be unambiguous.

\vskip0.5in

We acknowledge valuable discussions with C.P. Korthals Altes
and M. Teper.
This work was supported by PPARC grants GR/J21354 and
GR/K55752, and S.T.W. acknowledges the award of a PPARC studentship.

\end{document}